\documentstyle[aps,graphicx,prl,multicol]{revtex}
\input{epsf.tex}
\begin{document}
\draft
\title{Nature of the Fulde-Ferrell-Larkin-Ovchinnikov phases at low 
temperature in 2 dimensions}
\author{C. Mora and R. Combescot }
\address{Laboratoire de Physique Statistique,
 Ecole Normale Sup\'erieure*,
24 rue Lhomond, 75231 Paris Cedex 05, France}
\date{Received \today}
\maketitle

\begin{abstract}
We investigate, mostly analytically, the nature of the Fulde-Ferrell-
Larkin-Ovchinnikov superfluid phases for an isotropic two-dimensional 
superconductor, which is relevant for the experimental observation of 
these phases. We show that there is, in the low temperature range, a 
cascade of transitions toward order parameters with ever increasing 
complexity.
\end{abstract}

\begin{multicols}{2}

The transition to Fulde-Ferrel-Larkin-Ovchinnikov (FFLO) superfluid 
phases \cite{ff,larkov} has given rise to much work, but a full 
theoretical understanding of these inhomogeneous phases has not yet 
been reached. This is of high practical importance since experiment 
relies heavily on theory when it needs to test if these phases have 
actually been observed. Indeed beside their basic theoretical interest 
which, in addition to superconductivity and its coexistence with 
magnetism \cite{hbbm}, ranges from ultracold gases \cite{rcffan} to 
quark matter as it might occur in the core of neutron stars 
\cite{bowers,casal}, FFLO phases are also of very high applied 
interest, because they correspond to the superconducting phases which 
should appear in superconductors with very high critical fields in the 
vicinity of the transition to the normal state. A main strategy to observe 
these transitions in superconductors is to get rid of the orbital currents, 
responsible for the standard disappearance of the superconducting 
phase at low critical fields, and reach the regime of Pauli paramagnetic 
limitation. This can be done in clean quasi two-dimensional systems 
such as organic compounds made of widely separated conducting 
planes, or in high $T_{c}$  compounds. In these cases hopping 
between planes is very strongly inhibited. So when a strong magnetic 
field is applied parallel to the planes, the resulting orbital currents 
perpendicular to the planes are very weak and there is basically no 
orbital pair breaking effect which leaves open the possibility of 
transitions to FFLO phases at much higher fields. Nevertheless the 
small interplane coupling is necessary to suppress phase fluctuations 
and produce phase which can be properly described by a mean field 
theory. Indeed observations of this transition have been claimed quite 
recently in such organic compounds as $ \kappa $-(BEDT-
TTF)$_2$Cu(NCS)$_2$ \cite{nam,singleton} or $ \lambda $-(BETS-
TTF)$_2$GaCl$_4$ or FeCl$_4$ \cite{tanatar,uji}.

Actually when the field is not exactly parallel to the planes there are 
small orbital currents and associated vortices appear. The physics of the 
evolution from this vortex state to the pure FFLO phase is quite 
remarkable, both in the linear \cite{bul,shirai,manalo,hbbm} and in the 
non linear regime \cite{klrashi} since it involves a series of vortex states 
phases corresponding to Landau levels with increasing quantum 
number. Compared to this complex structure the limiting 2D FFLO 
phase sounds simple. Indeed the FFLO transition in 2D systems is 
believed to be second order and in particular Burkhardt and Rainer 
\cite{br} have studied in details the transition to a planar phase, where 
the order parameter $\Delta ( {\bf  r})$ is a simple $ \cos({\bf  q}. {\bf  
r})$ at the transition. This phase has been found by Larkin and 
Ovchinnikov \cite{larkov} to be the best one in 3D at $T=0$ for a 
second order phase transition. And in 3D it is also found to be the 
preferred one in the vicinity of the tricritical point and below 
\cite{matsuo,buz1,cm}, although in this case the transition turns out to 
be first order (except at very low temperature). However it is not clear 
that this simple order parameter is always the best one since, as first 
explored by Larkin and Ovchinnikov, it is in competition with any 
superposition of plane waves, provided that their wavevectors have all 
the same modulus.

In this paper we explore the low temperature range in 2D for the pure 
FFLO phase transition in the simple isotropic case and show that 
surprisingly there is a cascade of the second order transitions toward 
order parameters with ever increasing complexity. In principle there is 
an infinite number of phases. Therefore the pure FFLO phase display a 
structure just as rich as the vortex phases that we mentionned above, 
and the interplay between both promises to be of exceptional interest. A 
first step in this direction has been made recently by Shimahara 
\cite{shima2} who, instead of simple cosine, found a transition toward 
a superposition of up to six plane waves when looking for more 
complex 'cristalline' structures. We find agreement with his results. 
However we will see that cristallinity does not give the proper 
understanding of these phases. Instead we show that, when the 
temperature is lowered toward $T=0$, the cascade of transitions toward 
order parameters with an ever increasing number of plane waves is due 
to the singular nature of the $T=0$ limit, which is already appearent in 
the $T=0$ wavevector dependence of the second order term in the free 
energy expansion in powers of the order parameter.

Our study follows the general framework set up by Larkin and 
Ovchinnikov \cite{larkov} and used by Shimahara \cite{shima2} in the 
2D case. The free energy difference $\Omega \equiv  \Omega _{s} - 
\Omega _{n}$ between the superconducting and the normal state is 
expanded to fourth order in powers of the order parameter $\Delta ({\bf  
r})  = \sum  \Delta _{{\bf  q}_{i}} \exp(i{\bf  q}_{i}. {\bf  r})$. The 
second order contribution to $\Omega$ is $ N _{0}  \sum_{{\bf  q}} 
\Omega _{2}(q, \bar{ \mu },T) | \Delta _{{\bf  q}} |^{2}$, where $ N 
_{0}$ is the single spin density of states at the Fermi surface and $ 
\bar{ \mu } = (\mu _{\uparrow} - \mu_{\downarrow})/2 $ is half the 
chemical potential difference between the two spin populations. At the 
FFLO transition we have $\Omega _{2}(q, \bar{ \mu },T)=0$. 
Naturally the actual transition corresponds to the largest possible $ \bar{ 
\mu }$ at fixed $T$. It is easily seen \cite{larkov} that the 
corresponding wavevector $q$ for the order parameter is obtained by 
minimizing:
\begin{eqnarray}
I(\bar{q}, \bar{ \mu },T) = 2 \pi  T  \: {\rm Im} 
\sum_{n=0} ^{ \infty}  \frac{1}
{\sqrt{( i \bar{\omega  } _{n})^{2}-\bar{ \mu }^{2}  \bar{q}^{2}}} -
 \frac{1}{i \bar{\omega  } _{n}}
\label{eq1}
\end{eqnarray}
with respect to $\bar{q}$, where we have introduced the dimensionless 
wavevector $ \bar{q} = q v_{F} / 2 \bar{ \mu } $ and $\bar{\omega  
}_{n} \equiv \omega _{n} - i \bar{\mu }$ with $ \omega _{n} = \pi T 
(2n+1) $ being the Matsubara frequency. This can be rewritten as an 
integral over real frequency. At  $T=0$ one finds explicitely $ 
I(\bar{q}, \bar{ \mu },T) = {\rm Re} \ln ( 1+ \sqrt{1-\bar{q}^{2}}) - 
\ln 2$, which is minimum for $ \bar{q} = 1 $ , in agreement with 
Shimahara \cite{shima1} and Burkhardt and Rainer \cite{br}. We note 
that the location $ \bar{q} = 1 $ of the minimum is a singular point for 
$ I $ since $ I = \ln(\bar{q}/2)$ for $ \bar{q} > 1 $, so $ I $ is 
continuous while its derivative is discontinuous for $ \bar{q} = 1 $. 
For $ T \neq 0$ but low enough temperature, the condition giving the 
minimum can be expanded in powers of $t ^{1/2}$ where $ t = T /  
\bar{\mu}$ is a reduced temperature. We have found that expanding 
$\bar{q}-1$ up to $t^{2}$ gives a result in excellent agreement with a 
straight numerical evaluation for $ t < 0.1$. Here (and similarly below) 
we do not give the somewhat lengthy full expression \cite{longpaper}, 
but we omit the last two terms and display only the lowest order 
correction which is:
\begin{eqnarray}
\bar{q}-1 = \frac{t}{2} \,  \ln  \frac{\pi }{2t} 
\label{leading2}
\end{eqnarray}

The second order term in $\Omega$ fixes only the modulus, given by 
Eq.(2), of the $N$ wavevectors entering the plane wave expansion of 
the order parameter. This leaves open all the various possible order 
parameters arising from any combinations of these plane waves. The 
selected state will correspond to the lowest fourth order term, which is 
given \cite{larkov} in general by:
\begin{eqnarray}
\frac{ N _{0}}{2} \sum_{i,j} (2- \delta _{{\bf  q}_{i},{\bf  q}_{j}}) |  
\Delta _{{\bf  q}_{i}}| ^{2}  |  \Delta _{{\bf  q}_{j}}| ^{2} J(\alpha 
_{{\bf  q}_{i},{\bf  q}_{j}})  &  & \\ \nonumber 
+ (1-\delta _{{\bf  q}_{i},{\bf  q}_{j}}-\delta _{{\bf  q}_{i},-{\bf  
q}_{j}})  \Delta  _{{\bf  q}_{i}}  \Delta _{-{\bf  q}_{i}}  \Delta ^{*} 
_{{\bf  q}_{j}}  \Delta  ^{*}_{-{\bf  q}_{j}}  \tilde{J} (\alpha _{{\bf  
q}_{i},{\bf  q}_{j}})
\label{fourthorder}
\end{eqnarray}
where $\alpha _{{\bf  q}_{i},{\bf  q}_{j}}$ is the angle between ${\bf  
q}_{i}$ and ${\bf  q}_{j}$. Just as for the second order term $ 
J(\alpha)$ and $ \tilde{J} (\alpha)$ can be expressed \cite{longpaper} 
as frequency integrals, expanded at low temperature $ t \ll 1$ and 
evaluated for the optimal value of  $ \bar{q}$ (given by Eq.(2) to 
leading order). As for the optimal $ \bar{q}$ we have carried this 
expansion up to second order in $t ^{1/2}$. If we restrict ourselves to 
leading order as in Eq.(2) we find for example for $ J(\alpha)$:
\begin{eqnarray}
16 t \bar{ \mu }^{2} J(\alpha) = \frac{\pi }{\alpha } \frac{1}
{\cosh ^{2}[(1/4) \ln  \frac{\pi }{2t} - \beta ^{2} /2]}  &  & \\ 
\nonumber 
- \frac{8}{\sqrt{ \pi}(1+ \cos ( \alpha /2)) } 
\int_{0} ^{ \infty} \! dv  \: \frac{\exp(-v ^{2})}{ v ^{2}+\beta ^{2}} 
\label{Jalpha}
\end{eqnarray}
where we have set $ \beta ^{2} = (1-\cos ( \alpha /2)) /t $. The result up 
to second order in $t ^{1/2}$ is not much more complicated. A similar 
result \cite{longpaper} is found for $ \tilde{J} (\alpha)$.
\begin{figure}[h]
\begin{center}
\includegraphics[width=0.48\textwidth,height=0.3
\textheight]{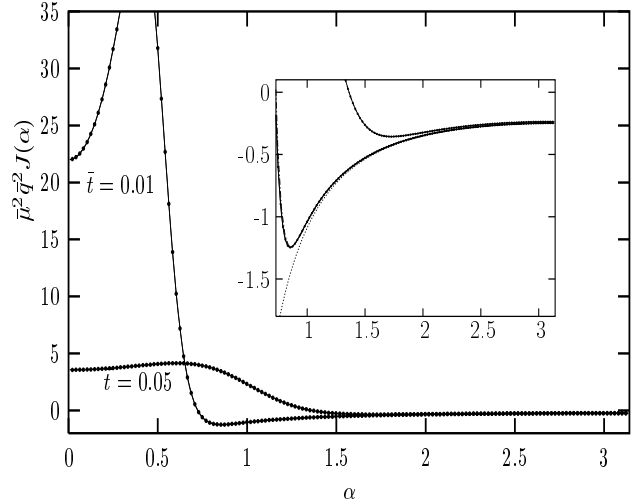}
\caption{$J(\alpha)$ for values $0.01$ and $0.05$ of
the reduced temperature $\bar{t} = T/(\bar{q} \bar{\mu}) =
t / \bar{q}$
(with essentially $ \bar{q}\approx 1$ in all this
range). The lines are calculated using our low temperature
expansion for $J(\alpha)$, while
the points correspond to the direct numerical summation over Matsubara
frequencies. The dashed line in the insert (which displays the same
results with different scale) is the  asymptotic behaviour for
$J(\alpha)$ in the $T \to 0$ limit Eq.(\ref{Jlargealpha}). }\label{fig1}
\end{center}
\end{figure}
\vspace{-8mm}
Plots of $ J(\alpha)$ for various low temperatures can be seen in Fig. 
\ref{fig1}, either obtained from these low temperature expressions or 
from exact calculations of $J(\alpha)$ obtained by direct numerical 
summation over Matsubara frequencies after analytical angular 
averaging.
We see that, at this level of accuracy, both agree remarkably well up to 
rather high temperatures which shows that our low temperature 
expansion is quite under control. Now Fig.1 shows that $ J(\alpha)$ 
has at low $T$ a quite remarkably structured behaviour which can be 
easily understood from some limiting cases. First let us take the limit $ 
T \rightarrow 0$ at fixed $ \alpha $. This implies $ \beta ^{2} 
\rightarrow \infty$, the first term in Eq.(4) goes to zero and the easy 
integration leads to:
\begin{eqnarray}
J(\alpha) = -  \frac{1 }{4  \bar{ \mu } ^{2}} \frac{1}{\sin^{2} ( \alpha 
/2)} 
\label{Jlargealpha}
\end{eqnarray}
On the other hand if at fixed $ T $ we take the limit $ \alpha \rightarrow 
0$, we have $ \beta ^{2} \rightarrow 0$. The limiting behaviour of the 
integral is $ \pi /(2 \beta ) - \pi ^{1/2}$, the dominant divergent 
contributions from the two terms in Eq.(4) cancel out giving:
\begin{eqnarray}
J(0) =  \frac{1 }{4  \bar{ \mu } ^{2}} \frac{1}{t} 
\label{eq23}
\end{eqnarray}
which goes naturally to infinity for $ T \rightarrow 0$. From these two 
limiting situations we can understand that, for most of the $ \alpha $ 
range$, J(\alpha)$ is negative as it can be seen from Eq.(5) and it goes 
to large negative values when $ \alpha $ gets very small. On the other 
hand for $ \alpha =0$ or very small $ J(\alpha)$ is positive and very 
large, as results from Eq.(6) (surprisingly $ J(\alpha)$ starts first to 
increase strongly before going down to very negative values).

We have made the same kind of treatment \cite{longpaper} for 
$\tilde{J} (\alpha )$. The limit $ T \rightarrow 0$ with fixed $ \alpha $ 
leads to $ \bar{ \mu } ^{2} \tilde{J}(\alpha) = -  \pi  / 4  \sin \alpha $ 
while in the limit $ \alpha \rightarrow 0$ at fixed $ T $ we find $  \bar{ 
\mu } ^{2} \tilde{J}(\alpha) =  - 1 /4 $. These two cases make 
reasonable that $\tilde{J} (\alpha)$ is always negative, as we find. 
However, just as for $J( \alpha )$, one finds also a singular behaviour 
at small $ \alpha $. While for $ \beta \approx  \alpha /(8t) ^{1/2} \gg 1$ 
and $\bar{ \mu } ^{2} \tilde{J} (\alpha)$ diverges as $ - \pi /4  \alpha 
$, it goes to the finite value $  - 1/4$ for $ \alpha =0$. Nevertheless the 
divergent behaviour in $ \alpha ^{-1}$ is weaker than the one in $ 
\alpha ^{-2}$ found for $J( \alpha )$ and similarly $J(0) \gg 
|\tilde{J}(0) |$ at low $T$. So $J( \alpha )$ will play the dominant role 
and at first we omit $\tilde{J} (\alpha)$ from our considerations.

Let us now consider which is the most favorable order parameter. We 
will always find that the fourth order term is positive, so we have to 
make it as small as possible in order to minimize the free energy. We 
note first that there is unevitably a strongly positive contribution, 
proportional to $J(0)$, coming from the N terms ${\bf  q}_{i}= {\bf  
q}_{j}$ in the first sum of Eq.(3). However their unfavorable effect is 
relatively reduced when $N$ increases since there are $ N ^{2}$ terms 
in this sum. Since we clearly have to avoid other positive contributions 
of this kind, we want to forbid the angle domain $ 0 < \alpha < 
\alpha_0$ where we can take $\alpha_0$ as giving $J(\alpha_0) = 0$, 
which is very close to the (strongly negative) minimum of  $J( \alpha )$ 
as it can be seen on Fig. 1. So the angle between any two different 
wavectors ${\bf  q}_{i}$ and $ {\bf  q}_{j}$ should be at least 
$\alpha_0$. On the other hand it is of interest  to have angles close to 
$\alpha_0$ since they give strongly negative contributions to $ \Omega 
$. Since it is better to have the number of wavevectors $N$ as large as 
possible, we come from symmetry reasons to the conclusion that we 
have to choose the wavevectors ${\bf  q}_{i}$ angularly equally 
spaced with the angle between two nearest wavevectors as close as 
possible to $\alpha_0$. Now $\alpha_0 \rightarrow 0 $ when $T 
\rightarrow 0 $. Therefore we have indeed a cascade of transitions 
corresponding to order parameters with an ever increasing number $ N$ 
of wavevectors when $T \rightarrow 0 $. This singular behaviour is 
clearly linked to the fact that it is not possible, in particular for 
$J(\alpha)$, to perform an expansion in the limit  $T \rightarrow 0 $, 
i.e. this limit is singular.

More precisely it is quite a difficult problem to handle this minimization 
from scratch, and for example to prove that the minimum $ \Omega $ is 
not obtained for a completely disordered situations. However symmetry 
makes such a result quite unlikely. On the other hand if we take for 
granted that the wavevectors are equally spaced angularly it is easy to 
prove that the weight $ | \Delta _{{\bf  q}_{i}} |$ are all equal, because 
this problem can be mapped on a tight binding Hamiltonian on a ring. 
Conversely if we assume that all these weights are equal the problem is 
equivalent to find the equilibrium position of atoms on a ring with 
repulsive short range interaction and attractive long range interaction. 
We expect the equilibrium to be a crystalline structure corresponding in 
our case to equally spaced wavevectors.
\vspace{-4mm}
\begin{figure}[h]
\begin{center}
\includegraphics[width=0.48\textwidth,height=0.3
\textheight]{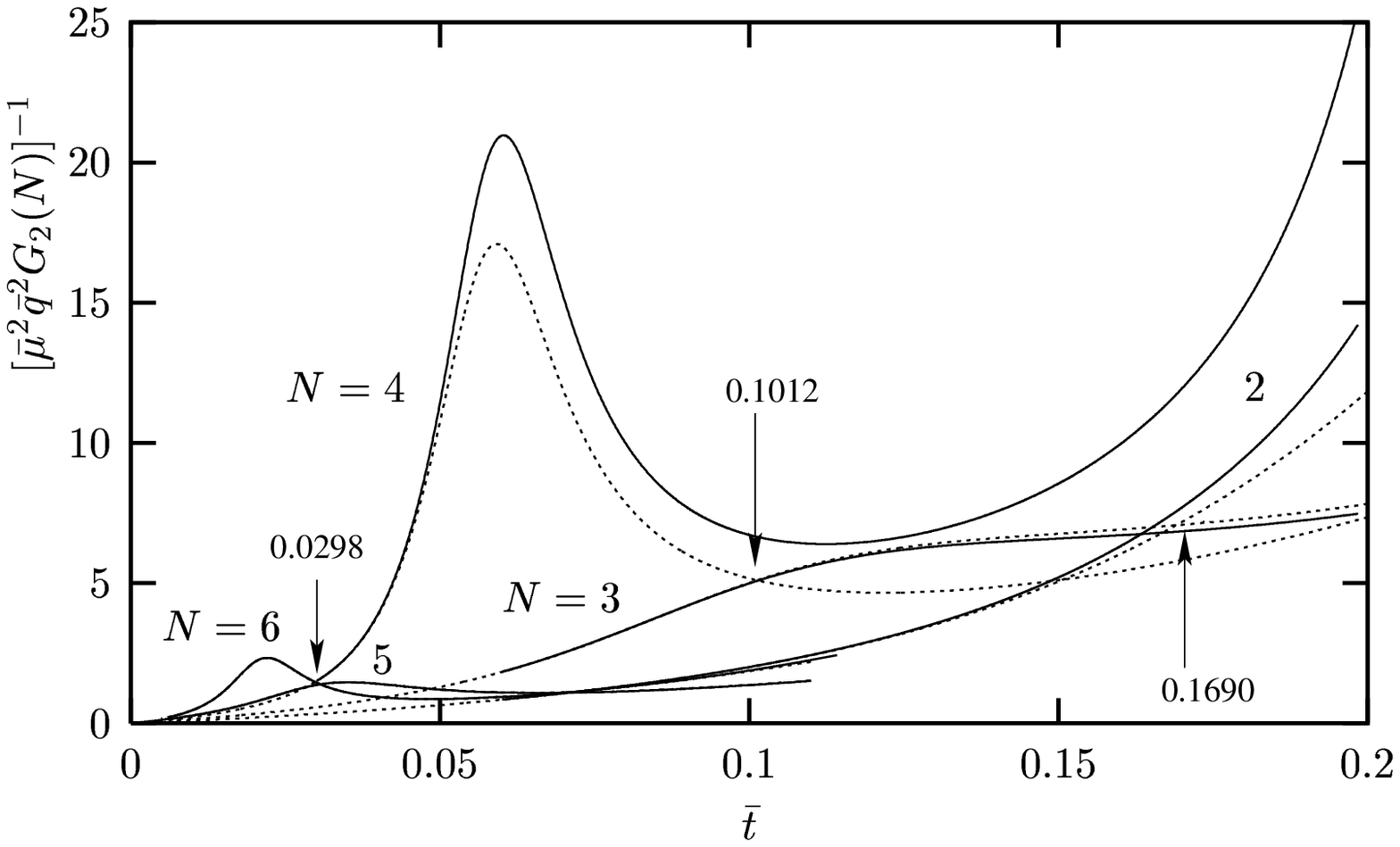}
\includegraphics[width=0.48\textwidth,height=0.3
\textheight]{alter-cascade-fin.eps2}
\caption{G$_2 ^{-1}$ as a function of temperature for various plane 
waves number $N$ (for clarity only solutions up to $N=14$ are 
presented). The arrows give the values of the successive transitions. In 
the upper panel the dashed lines are the exact Matsubara summations. 
The full lines in the upper panel, as well as all the lines in the lower 
panel, are from the low temperature expansion for $J(\alpha)$ and 
$\tilde{J} (\alpha )$. In the insert the full dots are the exact temperature 
locations of the transitions together with the corresponding values of 
$N$ (the empty dots just give $N-2$). The full line gives the low 
temperature evaluation of these transitions from the solution of a simple 
transcendental equation, and the almost undistinguishable dashed line 
comes from the explicit first iteration solution of this 
equation.}\label{fig3}
\end{center}
\end{figure}
\vspace{-8mm}
In the low temperature limit ($t \ll 1$), we can easily be more 
quantitative and write simple analytical answers for our problem. 
Indeed the result is dominated by the small $ \alpha $ behaviour of 
$J(\alpha)$. In this range Eq.(4) becomes:
\begin{equation}\label{asympto}
\bar{\mu}^2 J(\alpha) \simeq
\frac{\pi}{16 t \alpha \cosh^{2} X} - \frac{1}{\alpha^2}
\end{equation}
where $X= x - (1/4) \ln( \pi /2t)$ with $x=\alpha ^{2}/16 t$. The 
second term is the asymptotic form Eq.(\ref{Jlargealpha}). The zero of  
$J(\alpha)$ we are looking for is found for large $X$ and satisfies 
$e^{x} =(k/ \sqrt{2}) x^{1/4} $ with $k = (2 \pi ^{3}/ t ^{2}) 
^{1/4}$, which is easily solved iteratively for large $k$. The leading 
order $ x = \ln (k / \sqrt{2}) $ gives $ \alpha_0 = ( 8t \ln [(\pi 
^{3}/2)^{1/2}/t] ) ^{1/2}$. The first order iteration is $ x = \ln (k / 
\sqrt{2}) + 1/4 \ln (\ln (k / \sqrt{2}) $. Then the optimum value of $N$ 
is given by the integer value of $ 2 \pi /\alpha_0$, which conversely 
fixes the temperature at which the system switches from $N-2$ to $N$ 
wavevectors. In the insert of Fig.3, one sees that the first iteration gives 
results almost undistinguishable from the exact solution of the above 
transcendental equation for $x$.

Exact results can be obtained by handling numerically the various steps. 
The minimization of the free energy with respect to $ | \Delta _{{\bf  
q}} |$ gives $\Omega / N_0 = - \Omega_2 ^2 (q,\bar{\mu},T) / 
G_2(N)$ with:
\begin{equation}
\label{G2odd}
N G_2(N) = 2 J(0) + 4 \sum_{n =1} ^{N-1}  J(2 \pi n / N)
\end{equation}
where we have omitted the $\tilde{J}$ terms for simplicity. One sees in 
Fig. 3 the results (including the $\tilde{J}$ terms) for $ G_2^{-1} 
(N)$, i.e. essentially the free energy, for various values of $N$. The 
stable state has the highest  $ G_2^{-1} (N)$. The cascade of phase 
transitions is clearly seen in the lower panel, corresponding to the lower 
temperature range. The arrows show the successive transition 
temperatures. The results from straight numerical Matsubara summation 
can not be distinguished from those obtained from our low $T$ 
analytical expression for $J(\alpha)$ and $\tilde{J}(\alpha)$, except for 
the higher temperature, as seen in the upper panel. The critical 
temperatures for switching from $N-2$ to $N$ are shown as dots in the 
insert of Fig. 3. It can be seen that our above asymptotic calculation for 
these temperatures, shown as full curves, are actually quite good.

At low temperature we can make an asymptotic calculation of $ G_2 
(N)$. As seen on Fig. 1 $J(\alpha)$ switches rapidly above $\alpha_0 $ 
to its large angle asymptotic behaviour. Moreover at low $T$ the 
number of plane waves N is large and in Eq.(\ref{G2odd}) the sum is 
dominated by the small angles terms, which leads \cite{longpaper} to:
\begin{eqnarray}\label{G21}
\bar{\mu}^2 G_2(N) &\simeq& \frac{1}{2 N t} - \frac{N}{3}
\label{G22}
\end{eqnarray}
This expression corresponds in Fig. 3 to the rising part, on the low 
temperature side, of $ G_2^{-1} (N)$. On the other hand the downturn 
is due to contributions from the positive part of $J(\alpha)$ which can 
not be evaluated so easily. When we substitute in Eq.(\ref{G22}) the 
value $N = 2 \pi / \alpha _0$ for the optimum plane wave number we 
have: 
\begin{equation}
\bar{\mu}^2 G_2(N) = \frac{\alpha_0}{4 \pi t} \left( 1-
\frac{8 \pi^2 t}{3 \alpha_0^2} \right)
\end{equation}
Since $ 8t / \alpha_0^2 \sim 1 / \ln(1/t)$ from our above evaluation, we 
find that $ G_2(N)$ is always be positive in the low temperature range, 
which means that the transition stays always second order.
When we take into account the contribution of the $\tilde{J}$ terms in 
Eq. 3 that we had omitted so far, one gets an additional $- \ln N$ 
contribution in Eq.(\ref{G21}), which arises only when $N$ is even 
because for equally spaced wavevectors, the $\tilde{J}$ terms 
contribute only in this case since they require opposite wavevectors. 
This contribution is just enough to systematically tilt the balance in 
favor of even $N$. In this respect the phase $N=3$ found by 
Shimahara \cite{shima2} appears as an exception. Otherwise our 
cascade of phase transitions contains only order parameters with an 
even number of plane waves, and the order parameter is just a real sum 
of cosines, with still a degeneracy due to the freedom of choosing the 
phases in these cosines. 

We note that, although the transitions between the various FFLO phases 
and the normal phase are second order, we expect the transitions 
between different FFLO phases in the superfluid domain to be first 
order since one can not go continuously from a given order parameter to 
the next one. Finally we have seen that these successive transitions are 
directly due to the singularity which occurs in 2 D at $T=0$. This 
singularity itself arises because the two Fermi circles corresponding to 
opposite spins come just in contact when one applies the shift 
corresponding to the wavevector $ {\bf  q}$ of the FFLO phase. 
Therefore this singularity is a general feature of 2D physics and we 
expect it to give rise to similar consequences in more realistic and more 
complex models describing actual physical systems.

* Laboratoire associ\'e au Centre National
de la Recherche Scientifique et aux Universit\'es Paris 6 et Paris 7.

\end{multicols}
\end{document}